# Cognitive image processing: the time is right to recognize that the world does not rest more on turtles and elephants

Emanuel Diamant
VIDIA-mant, Israel
<emanl.245@gmail.com>
http://www.vidia-mant.info

**Abstract:** Traditional image processing is a field of science and technology developed to facilitate human-centered image management. But today, when huge volumes of visual data inundate our surroundings (due to the explosive growth of image-capturing devices, proliferation of Internet communication means and video sharing services over the World Wide Web), human-centered handling of Big-data flows is impossible anymore. Therefore, it has to be replaced with a machine (computer) supported counterpart. Of course, such an artificial counterpart must be equipped with some cognitive abilities, usually characteristic for a human being. Indeed, in the past decade, a new computer design trend – Cognitive Computer development – is become visible. Cognitive image processing definitely will be one of its main duties. It must be specially mentioned that this trend is a particular case of a much more general movement – the transition from a "computational data-processing paradigm" to a "cognitive information-processing paradigm", which affects today many fields of science, technology, and engineering. This transition is a blessed novelty, but its success is hampered by the lack of a clear delimitation between the notion of data and the notion of information. Elaborating the case of cognitive image processing, the paper intends to clarify these important research issues.

## 1. Introduction

Among the five human senses through which we explore and perceive our surrounding, vision takes a unique and remarkable place. A great part of our brain is permanently busy with processing the endless incoming video stream. To what extent is this phenomenon specific to our brain? – It is still an open question: Some think that visual data processing occupies about a half of our cortex [1], while other state "that 80% of the brain is used for processing visual data" [2], [3], [4].

This way or another, the central paradigm of vision, shaped by millions of years of natural evolution, has remained unchanged for the most part of our history: the light reflected or emitted from objects in the surrounding world enters the observer's eye and is focused onto a light-sensitive membrane in the back of the eye, called the retina. The retina serves as a transducer which converts the patterns of light into neuronal signals. Afterwards these signals (via the optic nerve) are carried to other parts of the brain for further processing [5].

The technological advances of the second millennium, particularly the invention and widespread of use of optical lenses, have greatly improved human vision abilities. New vision-enhancing and vision-strengthening implements (such as spectacles, microscopes and telescopes) have been put in use making it possible to extend human ability to see objects that are too small, too big, or too far away, to be perceived otherwise. Despite all these novelties, the traditional vision paradigm has remained unchanged – the image of the surrounding has been projected directly into the eye's retina.

For the first time, a paradigm shift in vision has come about only at the beginning of the nineteenth century. It was inspired by the invention of photography. In photography, image acquisition and its further examination by a human observer have been separated (in time and space) and for the first time were performed actually as two different and almost independent actions. Sooner or later, as a result of this separation, the notion of "image processing" has emerged and took its place as an inevitable part of a vision process. Wet development and fixation, retouching, brightness and contrast corrections were the first image processing procedures that have accompanied the advent of the photographic paradigm.

When a hundred and fifty years later the electronics era has begun, the "distributed vision" paradigm has not undergone significant changes. Only some of its pieces have been modified – instead of a plate- or a film-based camera an electronic camera was devised, instead of a chemical photosensitive emulsion a charge-coupled silicon sensor array was invented. Only the way in which an image was brought to observer's attention has been changed substantially – a novelty, called an electronic display, has been introduced and become ubiquitous. Consequently, the mode of image transfer from the acquisition camera to the image display has undergone impressive transformations – distantly located cameras and remote displays connected by various means of electronic communication are now a common arrangement for a large variety of installations: Closed Circuit Television, Space Mission Television, Underwater, Medical, Ultrasound, Tomographic and as such vision systems, which are now everywhere available. In the electronics era, rudimentary image processing which was characteristic for the early photographic age has evolved into a full-blown art, science and technology – in short, a widespread, common, and well established set of "best practice" processes.

Today there is a huge collection of excellent text-books and monographs covering the subject of present-day image processing. In the Reference list I provide only some of them – [6], [7], and [8] – with only one purpose in mind: not to endorse or to promote anyone of them, but simply to draw readers' attention to the huge amount of books and publications available now and targeting image processing techniques.

As follows from the above-mentioned literature, the vast number of image processing applications can be reduced to two main clusters: 1) image processing applications facilitating image access and delivery and 2) image processing applications facilitating human observer image perception. Image processing in the first group is concerned with image acquisition, image coding (for transmission, transfer, storage and retrieval purposes), and image display organization. Image processing in the second group is concerned with different aspects of image understanding and image analysis that always have been seen as a human observer prerogative.

However, the recent explosive growth of imaging technologies, the widespread availability of inexpensive video capturing devices, the proliferation of broadband Internet connections and the development of innovative video sharing services over the World Wide Web have led to an unprecedented flood of visual data (images and video) inundating our surrounding. Although visual data is not the only data circulating over the Internet, with the advent of the Big Data age, it has become the primary component of the today's data deluge.

"According to the International Data Corporation's recent report, "The Digital Universe in 2020," half of global big data — the valuable matter for analysis in the digital universe — was surveillance video in 2012, and the percentage is set to increase to 65 percent by 2015" writes Tiejun Huang in [9]. And continues: "Much like surveillance, the number of videos captured in classrooms, courts, and other site-specific cases is increasing quickly as well. This is the prelude to a "scene video" age in which most videos will be captured from specific scenes. In the near future, these pervasive cameras will cover all the spaces the human race is able to reach. In this new age, the 'scene' will become the bridge to connect video coding and computer vision research" [9].

It goes without further saying that the restricted human ability to cope and manage such massive volumes of visual data (images or "scene videos") is not appropriate anymore for this purpose and must be replaced by or delegated to a machine (a computer as we call it today). Self-evident, that such a computer has to possess a lot of human cognitive abilities which underpin understanding, analysis, and interpretation of images commonly attributed to a human observer.

Indeed, the next step in the Big Data management solution is declared to be the creation of the next generation computers equipped with Cognitive Computing capabilities. As [10] claims, this new computing paradigm, dubbed "Cognitive Computing", was proposed at first in 2011. Then, in 2013, IBM has declared Cognitive Computing to be the larger part of its future development program ('The Five in Five List"). In [11] it was claimed that IBM sees Cognitive Computing as its Next Disruptive Technology. Pursuing its goal to keep a leading position in Cognitive Computing R&D, IBM decided to combine its efforts with four other famous research institutions: Carnegie Mellon University, the Massachusetts Institute of Technology, New York University and Rensselaer Polytechnic Institute [12]. However, in the Cognitive Computing race, IBM and its allies are not alone – In 2011, Microsoft scientists presented research showing that cognitive computing techniques could be applied equally well to build computer systems aimed to understand human speech. Google's secretive X laboratory, known for inventing self-driving cars and augmented reality glasses, is also working several years on a simulation of the human brain [13].

Watching these impressive strides in the next generation computer development, it has to be mentioned that adjective "Cognitive" is not a stranger in the contemporary word-composition building practice. The adjective "cognitive" is now a respective component broadly used to create the majority of present-day (specification emphasizing) composite denominators. Vigilant readers are invited to test by themselves this tendency inquiring Google with an expression given in "double quote" marks and containing adjective "cognitive" as the expression's first component. For example, "Cognitive science" (returns 1,610,000 results), "Cognitive behavior" (407,000 results), "Cognitive computing" (331,000 results), "Cognitive biology" (22,200 results), and so on. (Please continue by yourself to explore further this issue). Recently, being invited to speak before people from the Endocrinology Department of the Sheba Hospital, I have put (just for curiosity's sake) the expression "Cognitive Endocrinology" as an inquiry for a Google search. To my great surprise Google has returned 630 hits (a valid confirmation of the expression's existence and use!)

The extensive use of the adjective "Cognitive" (in the new activity-specifying word-compositions) is not a quirk. It is a reflection of a growing understanding that any activity and any interaction of a living being with its surrounding is supervised and guided by its cognitive abilities. And this is true for all levels of life organization, from bacteria to Homo sapiens. Therefore, it is not surprising that when it comes to vision and image processing, the tendency that we witness is keeping on – Cognitive Vision and Cognitive Image Processing swiftly become an indispensable part of our vocabulary. ("Cognitive Vision" – 47,400 Google hits, "Cognitive Image Processing" – 65,700 Google hits). But these terms do not come as the latest fashion buzzwords – the terms "Cognitive Vision" and "Cognitive Image Processing" are coming to signify a critically important paradigm shift in vision and image processing (caused by the rapid progress in our technological means). And that is the reason that pushed me to further investigation into "Cognitive Image Processing" phenomena and to try to share with all of you my new findings.

My first impression gained from previous interactions with the subject was that the extensive use of the adjective "Cognitive" does not explain and does not tell nothing about its hidden meaning. And it seemed to me that it would be wrong to proceed further without paying attention to this strange unavailability. And, therefore, it would be wise (before we proceed further) to devote some more time and effort to explore the enigmatic nature of the term "Cognitive".

## 2. Cognitive: what does it really mean?

Bearing in mind that "cognitive" first of all assumes the human ability to think and reason, it seems quite natural to start to explore the issue by looking into the human related biological sciences like psychology, anthropology, neurosciences, and the like. Wikipedia tells us that the term 'cognition' dates back to the 15th century [14]. A lot of times later, the study of cognition has become a point of consideration in Europe and America. The recent new wave came about in the 1950s heralding the advent of the "Cognitive revolution" era and the birth of "Cognitive science" [15].

I do not intend to draw you into a review of the advancements in the evolution of the cognitive science – others would do this better than I (and more professional), for example [16], [17], and [18]. My goal was to find a suitable definition of the term "cognitive". And I did not succeed in this. But my failure is explainable: "After half a century of cognitive revolution we remain far from agreement about what cognition is and what cognition does. It was once thought that these questions could wait until the data were in. Today there is a mountain of data, but no way of making sense of it." [19].

The research community is sharply divided on almost any issue in cognitive discourse. Nevertheless, one proposition is generally accepted and approved unanimously: "cognition is a form of information processing and hence we can understand how the mind works and how organisms negotiate the world around them by understanding how information about the environment is represented, transformed, and exploited" [17]. And another supporting quotation: "Cognition is typically assumed to be information processing in a participant's or operator's mind or brain" [14]. And a lot of other examples like this one – "cognition is a form of information processing".

This assertion become part of cognitive science narrative in the late 1950's when the "Mathematical theory of communication" of Claude Shannon began to affect the studies of cognitive psychology. The shift was inspired by two famous publications – The George Miller's article "The magical number seven, plus or minus two: Some limits on our capacity for processing information" and Donald Broadbent's book "Perception and Communication". The powerful idea that has emerged from these works was the idea that we can understand how a cognitive system as a whole works by understanding how information flows through the system. Many psychologists and cognitive scientists subsequently took this type of information-processing flowchart to be a paradigm of how to explain cognitive abilities.

The advent of the computer era and the development of computational models of behavior that become a common practice at this time (the second half of the past century) have strengthened and accelerated the promotion of this view about "what cognition is and what cognition does", that is, promoting the idea that "cognition is information processing".

This general conviction is very good supported by Google's responses to such a "double quote marked" inquiry for "cognition is information processing" – 19,000 results in Google, 113 results in Scholar Google. A very impressive and undeniable endorsement! And we can be pleased with our exploration results, if only not a worrying remark – talking about information processing nobody does not say us what is on his mind, and what he means when the term "information" is being used. A short search in Wikipedia and other relevant sources quickly reveals that "information", generally taken for granted in cognitive and many other branches of science, does not have a consensus definition. And, despite widespread use of "information" as an established concept, no one actually knows what it means.

I have encountered this problem about ten years ago. And I have devoted a great part of my research to the study of the uncertainty of the information definition. Finally, I have invented a new, more suitable definition of information. And this will be the essence of my further discourse.

### 3. How to read this paper

The reader who reached this part of the paper certainly does not need to be reminded that the subject of our discourse – **cognitive image processing** – is a totally new idea that is just coming to a stage of its first blood feathers growing. Involvement in such a discourse is often a discomfiting challenge which usually requires a lot of patience and intellectual efforts. In this regard, I think, some tips and advices based on the author's personal experience (in such cases), will be certainly useful.

As it was said, the main goal of this paper is to promote the idea of cognitive image processing which in the nearest future will be our only means of coping with big-data (image) deluge.

Promoting new ideas has always been a risky business. The sad story of Giordano Bruno, who was burned alive by the Roman Catholic Church because he refused to abjure his ideas that were contradicting the official catholic doctrine, does not belong to the medieval history only. It's constantly reiterated through the whole human history.

Just about sixty years ago, in the Stalin's Soviet Union, whole fields of scholarship (physics, biology, genetics, cybernetics, linguistics, and political economy) were declared as ideologically incorrect and "bourgeois pseudo-sciences", while their proponents are being fired and dismissed from their jobs, arrested, sentenced to long-term imprisonment or to slave work in the corrective labor camps of Gulag.

Just about forty years ago, in the Mao Zedong's China, the Red Guards of the Cultural Revolution (the Hongweibings) have violently attacked their teachers and educators, their university professors, for being "capitalist intellectuals". Some of them were beaten to death, some were tortured by their students, some were humiliated and sent to a re-education in the "new life schools" (the labor concentration camps of Mao Zedong's China).

Today, in our liberal times, an attempt to promote a new idea will not throw you out in a forced re-education camp, but hostility, social isolation and mostly uninhabited surrounding is guaranteed for you. There is nothing to say about this situation, which is still a sad reality of our life. But, despite all of this, the life is going on and – don't worry… the history is repeating itself.

My first attempt to share with peers the humble results of my research has happened twelve years ago: my paper titled "Single pixel information content" was submitted to the 2nd Biologically Motivated Computer Vision workshop (BMCV 2002, Tubingen, Germany, November 22-24, 2002). The submission was strongly rejected. Reviewer's comments that were attached to explain the program committee's verdict sound like this: "The distinction between information and data processing is superficial – you have to be more specific (after all, data is information, isn't it?)".

I was then and am forever deprived the ability to argue with the program committee and to show it that the reviewer's conclusions are wrong. As a consequence, for more than twelve years, image processing community is preoccupied with its belief that information is data and data is information. Sure, that is not true, but they are doomed to a fatal delusion.

My last paper titled "Cognitive Surveillance" was submitted to the 11th IEEE International Conference on Advanced Video Signal-Based Surveillance (AVSS2014, August 26-29, 2014, Seoul, Korea), and was rejected too. The Assigned_Reviewer_4 has shaped his judgment in the following words: "The term "Cognitive Surveillance" is interesting to me. However, the author should cite more literature in the computer vision field to justify his arguments".

O, mine Gott! How can I explain to the educated reviewer that he is stumbled upon a totally new idea and the requested literature, which he hopes will help him to comprehend the matter, does not exist yet. And it will never come to existence, because Cognitive Surveillance has nothing to do with Computer Vision. It's a new idea, and you guys should work hard trying to catch its essence.

In this regard, the first advice that I would like to pass to my readers will be: Please, don't go the way my reviewers usually take on – looking for things they are already know and requiring the presence of things they are custom to deal with. Don't try to push me in the old shoes of your knowledge and understanding. Give me the right for my own bootstraps. (That is, try to follow me and do not attempt to reduce me to your level of understanding).

At the summer 2012, I was invited to write a chapter for the IGI Global's Encyclopedia of Information Science. The invitation was right on time – I already had a paper titled "When you talk about information processing what actually do you have in mind?" specially written for The 2012 Federated Conference on Computer Science and Information Systems (Wroclaw, Poland, 9-12 September 2012). It was, as usually, rejected and the IGI's invitation was a blessed opportunity to try again to get it published. I sent the manuscript and in a due time the reviewers' comments have arrived.

"The author does not provide significant reviews and background study in this paper… Information theory is not covered. Emphasis on personal definitions… This is a positioning article and has not to be included in this encyclopedia as it presents unpublished data in referred journals or scientific conference proceedings…"

O Good Lord! What could be said here? "You, Lord, who is rich in his mercy, who is gracious and compassionate, do not desert them (the semiliterate reviewers), with your unfailing love forgive them and reward everyone according to what they have done".

And to my readers I would like to say: Go to the Internet, ask Scholar Google, or the Research Gate, or the Cite Seer, or the TU Delft Repository. At least, inquire Google with double quotes around "Emanuel Diamant" – and you will be provided with an ample list of my publications (that the IGI's reviewer was unable to find by himself). After all, despite the reviewers' unfriendliness, from time to time I was lucky to publish something "in referred journals or scientific conference proceedings".

But again, I will insist on my advice to the readers: spare yourselves from reviewers' assistance, avoid publications that are certified by their approval. The best thing you can do – go straight to the arXiv repository! This is the most extensive repository which provides full texts of scholarly publications available free of charge and searchable for web engines. The arXiv is not peer reviewed, but a collection of moderators for each subject area glance through the manuscripts to ensure that the content is relevant and of interest to the current research field. Moderators' interventions are minor, – you can see by yourself: all of my once-rejected papers have been published in the arXiv Computer Science division.

The prime purpose of publishing a paper is to convey original results from the writer to the reader, to ignite a discussion, to exchange views and experience with others. The success in such an endeavor is measured by the number of readers who were attracted by the article, the number of article's downloads, and, finally, the number of citations that reference the publication. Scholar Google, Research Gate, Scopus and other search engines as a rule kindly provide such information about each paper's citation.

One day, examining with Scholar Google the list of my publications, I was surprised to find out that some of my old publications are escorted by a large number of citations. What's the matter? Why? Never before have been my papers favored with citations, certainly not with such amount. Further exploration has revealed that at some stage of my publication history trying to emphasize the continuity of my research I have referenced all previously written (and even not always published) articles in each of my new publication. In such a way, a bulk of self-references has arisen. In my eyes, that is certainly not a fair way to promote new ideas. So, the practice was quickly changed. In all my further publications, I stopped to refer to the relevant papers (pertinent to the current discourse) and instead of that I advise the interested readers to go to my site (http://www.vidia-mant.info) and to locate the needed paper by themselves, thus avoiding the accumulation of self-citations.

So, my next advice: do not await for a full definition of a new expression or a term in the body of a given paper. Go to the website and look there for a relevant paper where the matter has been once already explained. We both will benefit from this: I will be exempt from repeating again and again some basic components of the discourse and you will be always provided with an extended explanation of every crucial point encountered in the discourse.

And the last warning: Traditional (digital) image processing has evolved as a special technological means for the purposes of human-centered image management. There is a general agreement that efficiency and effectiveness of such a means will be greatly improved if its design will be based on some insights borrowed from the paradigm of natural vision. A rapidly growing movement known as "Biologically inspired design" has seized today different fields of engineering and design enterprises. Image processing, certainly, tries also to be a part of this movement. But in my view, incorporating biological analogies in an engineering design may be fruitful only if we have a good understanding about how the biological principles work in the nature. But that is usually not the case. Therefore, I strongly oppose the movement, and on several occasions have expressed my attitude towards it (My 2005 paper: "Does a plane imitate a bird? Does Computer Vision has to follow biological paradigms?" My 2009 paper: "Some considerations on how the human brain must be arranged in order to make its replication in a thinking machine possible". All this can be found on the web site).

Therefore, don't look for biologically inspired solutions neither in this manuscript, nor in the other ones. On the contrary, I am convinced that well-thought solutions developed for technological equivalents of biological functions could serve as insights to what's going on in their biological counterparts. In this way, interesting conclusions regarding various aspects of biological functions could be derived from their engineered imitations. Particular hypotheses about biological functions can be tested in such artificial replications given our inability to do such tests on real living systems. For that reason, my papers are always full with biology related associations and assumptions (You can call them "speculations", but one day the biological scientists will come across them and acknowledge their relevance).

At this time, my colleagues usually repel such digressions. European Conference on Computer Vision (ECCV 2006), rejecting one of my submissions, has stated this in a following way: "This is a philosophical paper... However, ECCV neither has the tradition nor the forum to present such papers. Sorry..."

O please! Don't apologize. Everyone has the right to freedom of thought, opinion and expression.

Taking into account all that was said just above, I would like to proceed with the further exploration of the **notion of cognitive image processing**. As you remember, we have stumbled on the enigmatic interrelationship between cognition and information. So, let us continue with our investigation.

## 4. Information: what is it and what is for?

We live in an Information Age and today the most commonly used word is "information". But what does it mean? What for is it being used? Nobody knows. And nobody doesn't feel any inconvenience about this. But in our case it is different.

Long time before, about thirty years ago, I was engaged in the homeland security research and development programme. Homeland security heavily relies on visual surveillance, and the latter, in turn, is busy with video streams acquisition and processing. While understanding what's going on in the acquired video was the prime goal of the whole enterprise, nobody never has not even approximated this goal, and therefore the mainstream efforts were directed only at image quality improvement and enhancement.

Aware of this bizarre situation, I have tried to approach the problem in a different way. Somehow, I came to a conclusion that what we are looking for in an image must be termed "image contained information". Attempts to comprehend what stands behind this term have taken another bundle of years. But at the year 2005, I have already published my first definition of information.

As it was already said, I have no intention to repeat again and again what has been many times expressed in my other papers (which are freely available on my web-site). But for the consistency of our discourse, I will provide a concise excerpt of the papers that represent my understanding of what information is (in general) and what is image information (in particular).

My definition of information sounds like this:
**"Information is a linguistic description of structures observable in a given data set".**

For the purposes of our discussion, digital image is a proper embodiment of such a data set. It is a two-dimensional array of a finite number of elements, each of which has a particular location and value. These elements are regarded to as picture elements (also known as pels or pixels). It is taken for granted that an image is not a random collection of these picture elements, but, as a rule, the pixels are naturally grouped into specific assemblies called pixel clusters or pixel structures. Pixels are grouped in these clusters due to the similarity in their physical properties (e.g., pixels' luminosity, color, brightness and as such). For that reason, I have proposed to call these structures **primary or physical data structures**.

In the eyes of an external observer, the primary data structures are further grouped and arranged into more larger and complex assemblies, which I propose to call **secondary data structures**. These secondary structures reflect human observer's view on the composition of the primary data structures, and therefore they **could be called meaningful or semantic data structures**. While formation of primary data structures is guided by objective (natural, physical) properties of data elements, ensuing formation of secondary structures is a subjective process guided by human habits and customs, mutual agreements and conventions between and among the observing group members.

As it was already stated above, **Description of structures observable in a data set should be called "Information".** Following the explained above subdivision of the structures discernible in a given image (in a given data set), two types of information must be distinguished – **Physical Information and Semantic Information**. They are both language-based descriptions; however, physical information can be described with a variety of languages (recall that mathematics is also a language), while semantic information can be described only by using natural human language.

I will drop the explanation how physical and semantic information are interrelated and interact among them. Although that is a very important topic, interested readers would have to go to the web-site and find there the relevant papers, which explain the topic in more details. Here, I will continue with an overview of the primary points that will facilitate our understanding of the issues of the further discourse.

Every information description is a top-down evolving coarse-to-fine hierarchy of descriptions representing various levels of description complexity (various levels of description details). Physical information hierarchy is located at the lowest level of the semantic hierarchy. The process of sensor data interpretation is reified as a process of physical information extraction from the input data, followed by an attempt to associate the input physical information with physical information already retained at the lowest level of a semantic hierarchy. If such association is achieved, the input physical information becomes related (via the physical information retained in the system) with a relevant linguistic term, with a word that places the physical information in the context of a phrase which provides the semantic interpretation of it. That is, the input physical information becomes named with an appropriate linguistic label and framed into a suitable linguistic phrase (and further – in a story, a tale, a narrative), which provides the desired meaning for the input physical information. (Again, more details can be found on the web-site).

The reader may experience some inconvenience caused by the use of new terminology often exploited in my texts. Don't worry, I am ready to provide you with a small pocket dictionary which will help you to overcome this difficulty. For example, the term "description" can be seen as an equivalent to the term "modeling" (widely used in other relevant publications). Physical structures delineation can be seen as a well-known to you image segmentation task. The coupling between physical and semantic information can be easily illustrated by a kinder recite: "Two dots, a comma, a circular trace, and here you have a human face."

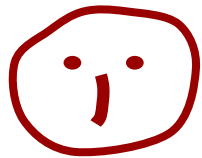

Here "Two dots, a comma, a circular trace" – are physical information descriptions, and the statement "here you have a human face" – that is a semantic information description. The latter is a label ("a human face") that is proposed to be our common denotation of the structure composed from the enumerated physical information components ("Two dots, a comma, a circular trace").

The segregation between physical and semantic information is the most essential insight about the nature of information. Its extraordinary importance should not be underestimated – the collapse of the Artificial Intelligence idea, the bankruptcy of the Machine Learning idea (semantic information is a mutual agreement between observers and thus cannot be learned, it must be only shared or granted), the failure of the everlasting attempts to derive semantic information from the omnipresent physical information (which is now a widespread practice in almost every scientific field) – all these are the result of the information dichotomy overlook.

Another important outcome from the semantic information definition is the comprehension about the form in which semantic information can be reified. That is, semantic information is always reified as a string of words, a piece of text, a narrative. That poses a problem when we think about further semantic information processing – computers are data processing machines, and how they can fit the task of text strings processing remains a problem and an unanswered question.

Extending further this line of thinking (about the consequences of semantic information notion uncovering and exploration), will promptly lead us to a departure from a currently dominating belief that "brain is a computational system" to a new and disrupting view that "brain is an information processing system". With an inevitable demolishing of such well-established and prosperous scientific fields as "computational biology", "computational behavioral", "computational neuroscience", "computational linguistics", "computational sociology", (and you can continue this list endlessly). Instead, newborn scientific fields are evolving – "cognitive biology", "cognitive behavioral", "cognitive neuroscience", "cognitive linguistics", "cognitive sociology", and so on.

To follow on this transition of the research paradigm is an amazing pursuit in itself. But it's without the scope of this paper. We have restricted ourselves to a more particular issue: Computer Vision research field passage into Cognitive Vision research field (while cognitive image processing is certainly an organic part of this passage). So, let us proceed with our duty.

But before we continue, I would like to reiterate the most important inferences that should be drawn from what was said until now: Cognitive information processing stands for semantic information processing, and semantic information processing stands for text processing. Lotfi Zadeh has proposed a special term for this: "Computing With Words" (more about this on my site, the paper "When you talk about information processing…"). However, Lotfi Zadeh does not know about semantic information and therefore his reasoning is inspired by an intuitive appreciation of physical information. My understanding about "what is semantic information processing" at this stage is also very limited and preliminary. And that is what I want to repeat again and again: I do know what is "semantic information"; I don't know what is "semantic information processing"! That is, at this stage I pretty well know what is semantic information, but how exactly it should be attributed to an image object is not so clear for me yet.

When I say I do not know yet how to attribute semantic information to a semantic object and then further use it for decision making and action planning in a frame of a text piece (a story, a narrative) – that is what "semantic information processing" really is – the emphasis is on "yet". The time is right to go and to investigate the subject urgently. There is no need to wait until the 1.39 billion euros dedicated to the Cognitive Robotics challenge would be wasted entirely. Or the budget of 1.19 billion euros for the Human Brain Project will be exhausted. Both are mega-projects launched by the European Commission, both are busy with brain information processing, but both do not ask the basic questions "What is cognition?", "What is information?", "What is information processing?". (For more on this, see a paper on the site: "Cognitive Robotics: for never was a story of more woe than this").

Bearing all these reservations in mind, let us proceed with our cognitive image processing subject.

## 5. The king is dead – long live the king!

The emergence of the "cognition-based" research paradigm and its antagonistic stance towards the "computation-based" paradigm is a long-lasting and widespread confrontation between the two approaches. But until now no advantage has been gained by either side of the confrontation. And the reason for this can be explained in three words: the battle grounds are undefined. The notions of data and information are rather empirical and intuitive, often used interchangeably and are followed by a heap of misleading claims and assumptions.

However, the new order which is established by the introduction of the up-to-date definition of information, especially introduction of the notion of information duality, puts an end to this endless mess. The victory of the cognition-based approach can be seen as a gateway to a higher level of processing, in our case: cognitive image processing. The crush of the "computation-based" paradigm is doubtless. Although the remains of its troops are still on the battlefield and are still alive. In such state of affairs, a post mortem analysis of these (refusing to give up) residues will be certainly useful and in place.

**Transform image processing.**
The classical image processing starts as a holistic approach. The notion of image information does not exist yet and its alliance with data structures (discernible in an image) is yet unknown. So, the image is accepted as a whole and therefore it is processed as a whole in the domain of image transforms. The most known transforms at that time are: The Fourier transform (including the Fast Fourier transform and the Discrete Fourier transform), The Wavelet transform, The Discrete Cosign transform. Despite great processing overheads – because the processing is done over the whole image – the techniques were popular on some stage of image processing history when the main purpose of processing was to facilitate human-centered image handling. Anyway, the image is finally offered for a human observer evaluation and what has happened with it beforehand does not matter.
When the intuitive understanding of the relevance of image structures has begin to emerge, transform image processing was abandoned in favor of structure considering processing approaches.
In the light of our today's understanding, transform image processing was a kind of image physical information description applied to a whole image and totally disregarding particular image structures.

In this vein, Shannon's information can be also seen as a kind of physical information description generalized over the whole image. Shannon's information is also an uncommon guest in the earlier times of image processing.

The utility of physical information that does not lead to further image content understanding (in the sense of semantic information unveiling) is fruitless and illusive.

**Object recognition, Image segmentation, and Edge detection.**

The growing intuitive understanding into the importance of image structure utilization has led to an extensive development of two closely related image processing techniques: image segmentation and edge detection. Their declared final goal is image objects detection and delineation. But at this stage of image processing evolution the notion of primary and secondary image structures does not exist yet, therefore definition "what is an image object" remains intuitive and vague. It is assumed that image object looks like a blob and it can be extracted from an image in two ways: first, by a border line that separates it from the rest of the image (that is called edge detection), or by partitioning the image into pixel patches corresponding to the supposed object blob (that is called image segmentation).

Both techniques are unconcerned about semantic roots of the object definition, and therefore both are trying to pursue their goals in a pure pixel-based bottom-up image processing fashion, (which can be associated with physical information processing, but cannot be seen as a task of semantic information delivery). They try to achieve the desired object delineation by means of physical information processing. But that is a wrong way to do image processing and that does not work!

**Image coding.**

Because effective image handling and management implies extensive image communication, image coding (for the purposes of image transmission and intermediate storage) has always been a very important topic in the hierarchy of image processing duties. The dominance of pixel-based image processing philosophy has its influence on image coding approaches too – both still image coding standard JPEG and moving pictures coding standard MPEG are implementing image coding techniques based on the pixel redundancy reduction. Object-based image content representation was always seen as the most relevant way of image coding (because human image perception is assumed to be done in a similar manner). The first attempt to utilize object-based image coding principles was made in MPEG-4 design requirements, but it was later abandoned because a suitable definition of image object is not attainable in a pixel-based processing environment.

The following versions of object-oriented image codes (MPEG-7 and MPEG-21) which have to inherit MPEG-4 coding techniques have also been blocked at some stage of their development. All further image coding progress (AVC, AIC, H.264 and JPEG standards) is implementing (in such or another way) the MPEG-2 pixel-block processing philosophy.

For the purpose of image delivery for further human observer evaluation (which is performed over the whole reconstructed (decoded) image) the intermediate image partition into geometric pixel blocks does not matter. That is the reason for the success and proliferation of the pixel-block coding technique utilized in the MPEG-2 and subsequent standards. But management of the expected Big data volumes will require a cognitive, that is, an object-based semantic information bearing image processing approaches, which contemporary MPEG image coding techniques would not be able to provide. Therefore JPEG/ MPEG coding techniques are not suitable for cognitive image processing and have to be urgently replaced.

**ROI – region of interest delineation and image salient zones revealing.**

Pixel-based image processing is a costly, energy and resources demanding task. To reduce the required expenses image processing designers seek to avoid entire image processing and restrict themselves only to a part of the image, to the so-called region of interest (ROI) processing. Without any doubts, delineation of such a region is a cognitive, that is, a semantic information processing task. But traditional image processing knows nothing about semantic information and its processing. Therefore traditional image processing tries to resolve the problem only in a particular way that it is familiar with – pixel-based physical information exploration. But physical information is not a suitable substitute for semantic information. The whole enterprise is a wrong idea.

The same argument is true for the task of visual saliency assessment, the saliency of some pixel group within an image. Such saliency assessment is usually conducted for an accomplishment of a ROI delineation. Again, pixel arrangements are physical information instances, salience is a semantic information instance – there is no way to get semantic information from a physical one. The problem has not a right solution, it is approached in a wrong way.

**CBIR – Content Based Image Retrieval.**
CBIR and the related methods – Content-based video retrieval (CBVR) and the combination of CBIR and CBVR Content-based visual information retrieval (CBVIR) method - are frameworks for efficiently retrieving images from image collections exploiting the similarity of the images. To judge about image similarity image content of the compared images has to be represented in some suitable form. At the early stages of CBIR systems design, the content of both images – the query image and the searched image in the collection – have been represented as a set of low-level features (feature vectors) derived from the processed images. The efficiency of such a search was low, therefore image content representations have shifted to semantic image content descriptors. That means, image objects represented by feature vectors were forced to be replaced with object semantic labels. That means, images have to be first of all annotated (manually or automatically). It turned out that the annotation process, sufficiently well being executed manually by humans, become an impassible hurdle when it comes to an automatic image annotation. The problem is known as the "semantic gap" barrier and further success in CBIR techniques implementation is crucially dependent on the success in the task of semantic gap bridging. As it follows from our earlier physical and semantic information notion definitions, this problem can't be resolved, because semantic information can't be derived from physical information (low-level feature vectors, without any doubt, are physical information descriptors and object labels are semantic information descriptors, and these are different information entities that cannot be mixed or interchanged).

Image processing technologies developed over the last 50 years are totally useless in the information processing age. But they are still in use and are even successful in some particular cases where semantic information is implicitly introduced into the image evaluation process. There may be always a particular solution suitable for a particular case, but the solution can never be generalized (because semantic information is always particular and subjective), and therefore the solution cannot be seen as a new paradigm or a paradigm shift.

## 6. Conclusions

The aim of this paper was to draw readers' attention to the very important paradigm shift in image processing practice pertinent to handle the ever growing amount of digital images inundating our surrounding. The traditional data-based (pixel-based) image processing paradigm is discharged of its duties and must to give up to a new cognitive (information-based) image processing paradigm. This is not an accidental case. This is a crucially important paradigm shift that presumes not only a change in image processing philosophy but a drastic change in our world perception and evaluation practice, which encompasses not only digital image processing, but the whole philosophy of living beings that act and interact with their surroundings.

Meanwhile, people are busy with gimmicks like "Cognitive Computing" overlooking that Cognitive Computing is an oxymoron (something that combines contradictive meanings, something like "exact approximation" or a "stupid wise man"). Because "Cognitive" implies information processing and "Computing" implies data processing, (and the two are incompatible). But who cares?